\documentclass[journal]{IEEEtran}

\ifCLASSINFOpdf
\usepackage[pdftex]{graphicx}
\else
\fi
\usepackage[toc,acronym]{glossaries}
\usepackage{adjustbox}

\newacronym{SPA}{SPA}{Single Processor Approach}
\newacronym{HPL}{HPL}{High Performance Linpack}
\newacronym{HPCG}{HPCG}{High Performance Conjugate Gradients}
\newacronym{HW}{HW}{Hardware}
\newacronym{SW}{SW}{Software}
\newacronym{OS}{OS}{Operating System}
\newacronym{PD}{PD}{Propagation  Delay}
\newacronym{PSO}{PSO}{Particle Swarm Optimization}
\newacronym{SoC}{SoC}{System-on-Chip}

\usepackage{tikz}
\usepackage{pgfplots}
\pgfplotsset{compat=1.14} 
\definecolor{webbrown}{rgb}{.6,0,0}
\definecolor{webyellow}{rgb}{0.98,0.92,0.73}
\definecolor{webgray}{rgb}{.753,.753,.753}
\definecolor{webblue}{rgb}{0,0,.8}
\definecolor{webgreen}{rgb}{0, 0.5, 0} % less intense green
\definecolor{webred}{rgb}{0.5, 0, 0}   % less intense red

\pgfplotscreateplotcyclelist{my color list}{%
	solid, color=webblue, every mark/.append style={solid, fill=webblue}, mark=*\\%
	densely dashdotted, color=webred, every mark/.append style={solid, fill=webred},mark=diamond*\\%
	densely dotted, color=webgreen, every mark/.append style={solid, fill=webgreen}, mark=triangle*\\%
	loosely dashed, color=webbrown, every mark/.append style={solid, fill=webbrown},mark=*\\%
	dotted, color=webyellow, every mark/.append style={solid, fill=webyellow}, mark=square*\\%
	densely dotted, every mark/.append style={solid, fill=gray}, mark=otimes*\\%
	dashed, every mark/.append style={solid, fill=gray},mark=diamond*\\%
	densely dashed, every mark/.append style={solid, fill=gray},mark=square*\\%
	dashdotted, every mark/.append style={solid, fill=gray},mark=otimes*\\%
	dasdotdotted, every mark/.append style={solid},mark=star\\%
}

\newcommand*{\boxcolor}{orange}
\makeatletter
\renewcommand{\boxed}[1]{\textcolor{\boxcolor}{%
		\tikz[baseline={([yshift=-1ex]current bounding box.center)}] \node [rectangle, minimum width=1ex,rounded corners,draw] {\normalcolor\m@th$\displaystyle#1$};}}
\makeatother

\begin{document}
	\title{Re-evaluating scaling methods\\ for distributed parallel systems}

	\author{J\'anos V\'egh% <-this % stops a space
		\thanks{Projects no. 125547 has been implemented with the support provided from the National Research, Development and Innovation Fund of Hungary, financed under the K funding scheme.
%			Also the ERC-ECAS support of project 886183 is acknowledged.
		}% <-this % stops a space
\thanks{Kalim\'anos BT, 4032 Debrecen, Hungary}% <-this % stops a space
\thanks{Manuscript submitted to IEEE transactions February 16, 2020; revised ??, 2020.}}

	%% The paper headers
	\markboth{Re-evaluating scaling methods for distributed parallel systems,~Vol.~14, No.~8, August~2020}%
	{J\'anos V\'egh: Re-evaluating scaling methods\\ for distributed parallel systems}
	%
	%
	%
	%
	%% If you want to put a publisher's ID mark on the page you can do it like
	%% this:
	%%\IEEEpubid{0000--0000/00\$00.00~\copyright~2015 IEEE}
	%% Remember, if you use this you must call \IEEEpubidadjcol in the second
	%% column for its text to clear the IEEEpubid mark.
	%
	%
	%
	%% use for special paper notices
	%%\IEEEspecialpapernotice{(Invited Paper)}
	\maketitle
	
	%% As a general rule, do not put math, special symbols or citations
	%% in the abstract or keywords.
	\begin{abstract}
		The paper explains why
		Amdahl's Law shall be interpreted specifically for distributed parallel systems
		and why it generated so many debates, discussions, and abuses.
		We set up a general model and list many of the terms affecting parallel processing. We scrutinize the validity of neglecting certain terms in different approximations, with special emphasis on the famous scaling laws of parallel processing. We clarify that when using the right interpretation of terms, Amdahl's Law is the governing law of all kinds of parallel processing.
		Amdahl's Law describes among others the history of supercomputing, 
		the inherent performance limitation of the different kinds of parallel processing
		and it is the basic Law of the 'modern computing' paradigm, 
		that the computing systems working under extreme computing 
		conditions are desperately needed.
	\end{abstract}
	%
	%
	%
	%
	%
	%% For peer review papers, you can put extra information on the cover
	%% page as needed:
	%% \ifCLASSOPTIONpeerreview
	%% \begin{center} \bfseries EDICS Category: 3-BBND \end{center}
	%% \fi
	%%
	%
	%
	%
	\section{Introduction}
	\IEEEPARstart{A}{mdahl} in his famous paper~\cite{AmdahlSingleProcessor67}, even in the title,
	wanted to draw the attention to that (as he has coined the wording) the  \gls{SPA}
	seriously shall limit the achievable computing performance, given that 
	"\textit{the organization of a single computer has reached its limits}" and attempted 
	to explain why it was so. Unfortunately,  his successors nearly completely misunderstood his intention. Rather than developing "\textit{interconnection of a
		multiplicity of computers in such a manner as to permit cooperative solution}"
	the followers used his idea only to derive the limitations of computing systems built from components manufactured for the \gls{SPA}. The successors constructed his famous formula as well, and unfortunately, attributed an inadequate meaning to its terms.  The quick technical development suppressed the real intention and meaning of the Law.
	When the computing needs and possibilities reached the point where the 
	the precise meaning of the Law matters, the incorrect interpretation attributed to its
	terms did not describe the experiences, giving way to different other 'laws' and scaling modes. With the proper interpretation, however, we show
	"\textit{that Amdahl's Law is one of the few fundamental laws of computing}"~\cite{AmdalsLaw-Paul2007}. Not only of computing, but of all --
	even computing unrelated -- partly parallelized otherwise sequential activities.
	This paper discusses only the consequences of the idea on scaling of computer systems constructed in \gls{SPA}; for the idea he was thinking about, see~\cite{IntroducingEMPA2018,VeghEMPAthY86:2016,VeghEMPA:2016}.
	
	In section \ref{sec:TheScaling}, the considered scaling methods
	are shortly reviewed, and some of their consequences discussed.
	Section \ref{sec:AmdahlsLaw},  describes Amdahl's idea shortly: and introduces his famous formula using our notations.
	In section \ref{sec:GustafsonLaw}, we scrutinize the basic idea of
	the massively parallel processing, Gustafson's idea.
	In section~\ref{sec:ModernLaw} we discuss the "modern scaling law"~\cite{VeghHowMany:2020}, based on the the recently introduced  "modern computing"~\cite{VeghModernParadigm:2019};
	essentially Amdahl's idea applied to the modern computing systems.
	
	Section \ref{sec:TheModel} introduces a (by intention strongly simplified and non-technical)
	model of the parallelized sequential processing.
	The model visualizes the meaning of the "parallelizable portion" and enables us to draw the region of validity of the "strong" and "weak" scaling methods.
	As those scaling methods and principles are relevant to all fields of parallel and distributed processing, we demonstrate the application of the presented formalism for different tasks in section~\ref{sec:ApplicationFields}.

	\section{The scaling methods}
	\label{sec:TheScaling}
	
	The scaling methods used in the field are necessarily approximations to the general model presented in section~\ref{sec:TheModel}. We discuss the nature and validity of those approximations, furthermore introduce
	the notations and the formalism.

	\subsection{Amdahl's Law\label{sec:AmdahlsLaw}}
	
	Amdahl's Law (called also the 'strong scaling') is usually formulated with an equation such as
	\vspace{-.3\baselineskip}	
	\begin{equation}
	S^{-1}=(1-\alpha) +\alpha/N \label{eq:AmdahlBase}
	\end{equation}
	
	\noindent where $N$ is the number of parallelized code fragments, 
	$\alpha$ is the ratio of the parallelizable fraction to the total (so $(1-\alpha)$ is the "serial percentage"),
	$S$ is a measurable speedup.
	That is, Amdahl's Law considers a \textit{fixed-size problem},
	and the $\alpha$ portion of the task is distributed to the fellow processors.

	When calculating the speedup, one calculates
	\begin{equation}
	S=\frac{(1-\alpha)+\alpha}{(1-\alpha)+\alpha/N} =\frac{N}{N\cdot(1-\alpha)+\alpha}
	\end{equation}
	However, as expressed in~\cite{AmdalVsGustafson96}: "\textit{Even though Amdahl's Law is theoretically correct, the serial percentage is not practically obtainable.}"
	That is, concerning $S$ there is no doubt that it corresponds to the ratio of
	\textit{the measured execution times}, for the non-parallelized and the parallelized case, respectively.
	But, what is the exact interpretation of $\alpha$, and how can it be used?

	Unfortunately, Amdahl used $\alpha$ with the meaning "\textit{the fraction of the
		the number of instructions which permit parallelism}" in Fig.~1 he used as an illustration in his paper. The illustration refers to the case when "\textit{around a point corresponding to 25\% data management overhead and 10\% of the problem operations
		forced to be sequential}". At that "point", there is no place to discuss
	more subtle details of the performance affecting factors (otherwise mentioned by Amdahl, such as "boundaries are likely to
	be irregular;
	interiors are inhomogenous;
	computations required may be dependent on the states
	of the variables at each point;
	\textit{propagation rates of different physical effects may be quite different};
	the rate of convergence or convergence at all may be strongly dependent on sweeping through the
	array along different axes on succeeding passes").
	It is worth to notice that Amdahl has foreseen issues with "sparse" calculations
	(or in general: \textit{the role of data transfer}); 
	furthermore that the \textit{physical size} of the computer and the interconnection of the computing units also matters. The latter is  crucial in the case of distributed systems.

	However, at that time (unlike today~\cite{PerformanceCounter2013,Molnar:2017:Meas}) the execution time was strictly determined by the number of the executed instructions.
	What he wanted to say was "\textit{the fraction of the time spent with executing the instructions which permit parallelism}" (at other places the correct expression "\textit{the fraction of the computational load}" was used). 
	This (unfortunately formulated) phrase "\textit{has caused nearly three decades of confusion in the parallel processing community. This confusion disappears when we use processing times in the formulations}"~\cite{AmdalVsGustafson96}.
	On one side, \textit{the researchers guessed that Amdahl's Law was valid only for software} (for the executed instructions), and on the other side \textit{the other affecting factors, he mentioned but did not discuss in details, were forgotten}.
	
	As expressed correctly in~\cite{AmdalVsGustafson96}: "\textit{For
		example, if the serial percentage is to be derived from computational experiments, i.e., recording the total parallel elapsed time and the parallel-only elapsed time, then it can contain all overheads, such as communication, synchronization, input/output and memory access. The law offers no help to separate these factors. On the other hand, if we obtain the serial percentage by counting the number of total serial and parallel instructions in a program, then all other overheads are excluded. However, in this case, the
		predicted speedup may never agree with the experiments.}"

	Really, one can express $\alpha$ from Eq.~(\ref{eq:AmdahlBase}) in terms measurable experimentally as
	
	\begin{equation}
	\alpha = \frac{N}{N-1}\frac{S-1}{S} \label{equ:alphaeff}
	\end{equation}
	
	\noindent
	That is this $\alpha_{eff}$ value, the \textit{effective parallel portion}, can be derived from the experimental data for 
	the individual cases.
	Also, it is useful to express the \textit{efficiency} with the pseudo-experimentally measurable
	\vspace{-.3\baselineskip}	
	\begin{equation}
	\boxed{E(\large N,\alpha)} = \frac{S}{N}=\boxed{\frac{1}{\textcolor{red}{\Large N}\cdot(1-\alpha)+\alpha}}= \frac{R_{Max}}{R_{Peak}}
	\label{eq:soverk}
	\end{equation}
	data, because for many parallelized sequential systems
	(including the TOP500 supercomputers) the efficiency (as $R_{Max}/R_{Peak}$) and the number of processors $N$ are provided. Reversing the relation, the value of $\alpha_{eff}$ can be calculated as
	\begin{equation}
	\boxed{\alpha(E,N)} = \boxed{\frac{E\cdot N -1}{E\cdot (N-1)}}\label{eq:alphafromr}
	\end{equation}
	
	As seen, \textit{the efficiency is a two-parameter function}
	(the corresponding surface is shown in Fig.~\ref{fig:EffDependence2018LogA}), demonstratively underpinning that "\textit{This decay in performance is not a fault of the
		architecture, but is dictated by the limited parallelism}"~\cite{ScalingParallel:1993}
	and that the properly interpreted Amdahl's Law perfectly describes its dependence on its variables.
	This proper interpretation also means that Amdahl's Law (after the pinpointing given in section~\ref{sec:ModernLaw}) shall describe the behavior of systems using a variety of parallelism, see section~\ref{sec:ApplicationFields}.

	\subsection{Gustafson's Law\label{sec:GustafsonLaw}}
	
	Partly because of the outstanding achievements of the parallelization
	technology, partly because of the issues around the practical utilization of Amdahl's Law, the 'weak scaling' (also called Gustafson's Law~\cite{Gustafson:1988}) was also introduced, meaning that \textit{the computing resources grow proportionally with the task size}. Its formulation was (using our notations)
	
	\begin{equation}
	S = (1-\alpha) + \alpha \cdot N \label{Equ:Gustafson}
	\end{equation}
	
	Similarly to the Amdahl's Law, the efficiency can be derived for the Gustafson's Law as (compare to Eq.~(\ref{eq:soverk}))
	%\noindent Notice that Amdahl's Law (as expressed by Eq.~(\ref{eq:AmdahlBase})) is \textit{independent} of
	%the number of processors, while Gustafson's Law (as expressed by Eq.~(\ref{Equ:Gustafson})) depends on it. Mathematically, Gustafson's formulation  cannot be \textit{directly} used to observe N's impact
	%on speedup since it contains an N dependent variable
	\begin{equation}
	\boxed{E(N,\alpha)} = \frac{S}{N} = \boxed{\alpha  + \frac{(1-\alpha)}{N}}  \label{Equ:GustafsonEff}
	\end{equation}
	
	\noindent From these equations immediately follows that
	the speedup (the parallelization gain) \textit{increases} linearly with the number of processors, \textit{without limitation}. This conclusion was launched amid much fanfare. They imply, however, some more immediate  conclusions, such as
	
	\begin{itemize}
		\item some speedup can be measured even if no processor is present
		\item the efficiency slightly \textit{increases} with the number of processors $N$ (the more processors, the better efficacy) 
		
		\item the non-parallelizable portion of the job either shrinks as the number of processors grows, or despite that, that portion is non-parallelizable,  the system distributes it between the $N$ processors 
		\item executing the extra instructions needed by the first processor to organize the joint work needs no time
		\item  all non-payload computing contributions such as
		communication (including network transfer), synchronization, input/output and memory access take no time
	\end{itemize}
	However, an error was made in deriving Eq.~(\ref{Equ:Gustafson}): \textit{the $N-1$ processors
		are idle waiting while the first one is executing the 
		sequential-only portion}. Because of this, the \textit{time} 
	that serves as the base for calculating the \textit{speed}up
	in the case of using $N$ processors 
	
	\begin{align*}
		T_{N} & =(1- \alpha)_{processing} + \alpha\cdot N \textcolor{red}{+ (1- \alpha)\cdot(N-1)_{idle}}\\
		& = (1- \alpha) \cdot N + \alpha\cdot N \\
		& = N\\
	\end{align*}
	For the meaning of the terms in~\cite{Gustafson:1988}, the author used the wording "is the amount of time spent (by a serial processor)".
	
	That is, before fixing the arithmetic error, impossible conclusions follow, after fixing it,
	the conceptual error comes to light: \textit{the weak scaling 
		assumes that the single-processor efficiency can be transferred to the parallelized sequential subsystems without loss}, i.e., that the efficacy of a system comprising $N$ single-thread processors remains the same than that of a single-thread processor. This statement strongly contradicts the experienced 'efficiency' of the parallelized systems,
	not speaking about the 'different efficiencies'~\cite{VeghHowMany:2020}, see also Fig.~\ref{fig:EffDependence2018LogA}.

	That is, the Gustafson's Law is simply a misinterpretation of the argument  $\alpha$: a simple function form transforms  Gustafson' Law to Amdahl's Law~\cite{AmdalVsGustafson96}. After making that transformation, the two (apparently very different) laws become identical. However, as suspected by~\cite{AmdalVsGustafson96}: "\textit{Gustafson's formulation gives an illusion that as if N can increase indefinitely}".
	This illusion led to the moon-shot of targeting to build
	supercomputers with computing performance well above the feasible (and reasonable) size and may lead to false conclusions in the case of using clouds. The 'real scaling' also
	explains why one could not reveal this illusion for decades
	and why it provoked decades-long debates in the community.

	\subsection{Real scaling\label{sec:ModernLaw}}
	
	The role of $\alpha$ was theoretically
	established~\cite{Karp:parallelperformance1990}
	and the phenomenon itself,
	that the efficiency (in contrast with Eq.~(\ref{Equ:GustafsonEff})) \textit{decreases} as the 
	number of the processing units \textit{increases},
	is known since decades~\cite{ScalingParallel:1993} (although it was not formulated in the functional form given by Eq.~(\ref{eq:soverk})).
	In the past decades,  the theory faded; mainly due to the quick development of the parallelization technology and single-processor performance.
	The community used the 'weak scaling' approximation 
	to calculate the expected performance values, 
	in many cases outside its range of validity. 
	The 'gold rush' for building exascale computers
	finally made obvious that under the extreme
	conditions represented by the need of millions of processors 
	the used 'weak scaling' leads to false conclusions: it "\textit{can be seen in our current situation where the historical ten-year cadence between the attainment of megaflops, teraflops, and petaflops has not been the case for exaflops}"~\cite{ExascaleGrandfatherHPC:2019}.
	It looks like, however, that in the feasibility studies of supercomputing using
	parallelized sequential systems  an analysis
	whether building computers of such size is feasible (and reasonable) remained out of sight either in USA~\cite{NSA_DOE_HPC_Report_2016,Scienceexascale:2018} or in  EU~\cite{EUActionPlan:2016}
	or in Japan~\cite{JapanExascale:2018} or in China~\cite{ChinaExascale:2018}.

\begin{figure}
		\includegraphics[width=\columnwidth]{%fig/
			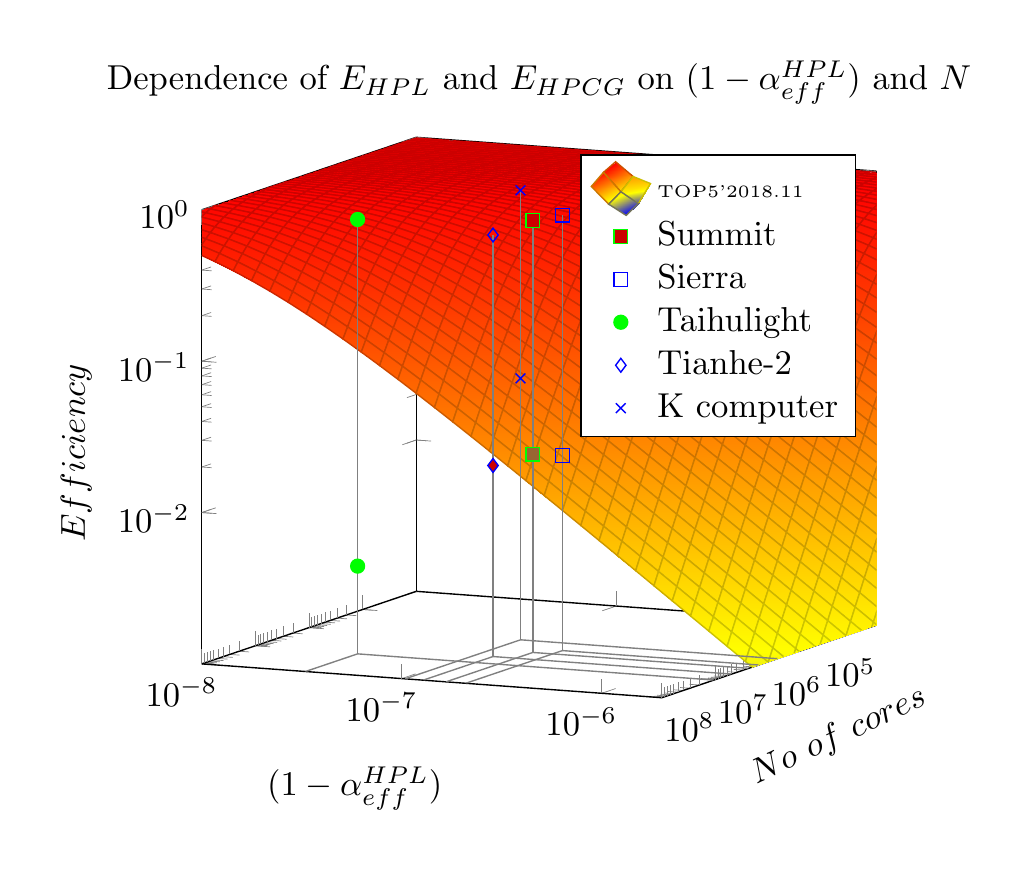}
		\caption{The 2-parameter efficiency surface (in function of the parallelization efficiency measured by benchmark \gls{HPL} and the number of the processing elements) as concluded from Amdahl's Law (see Eq.~(\ref{eq:soverk})), in the first order approximation. Some sample efficiency values for
			some selected supercomputers are shown, measured with benchmarks \gls{HPL} and \gls{HPCG}, respectively.  		\label{fig:EffDependence2018LogA}
		}
	\end{figure}

	Figure~\ref{fig:EffDependence2018LogA}
	depicts the two-parameter efficiency surface stemming out from Amdahl's law (see Eq.~(\ref{eq:soverk})). 
	On the surface, some measured efficiencies of the present top supercomputers are also depicted, just to illustrate some general rules. 
	To validate the model described in section~\ref{sec:TheModel}
	the data of the rigorously verified supercomputer database~\cite{TOP500:2017}
	was used, as described in~\cite{VeghHowMany:2020}.
	The \gls{HPL}\footnote{http://www.netlib.org/benchmark/hpl/} efficiencies are sitting on the surface, while
	the corresponding \gls{HPCG}\footnote{https://www.epcc.ed.ac.uk/blog/2015/07/30/hpcg} values are much below those values.
	The conclusion drawn here was that "\textit{the supercomputers have two different efficiencies}"~\cite{DifferentBenchmarks:2017},
	because one cannot explain the experience in the frame of 
	the "classic computing paradigm".

	%As Figure~\ref{fig:EffDependence2018LogA} witnesses,
	The $Taihulight$ and $K~computer$ stand out from the "millions core" middle group. Thanks to its 0.3M cores, $K~computer$ has the best efficiency for the $HPCG$ benchmark, while $Taihulight$ with its 10M cores the worst one.   
	The middle group follows the rules~\cite{VeghHowMany:2020}. For $HPL$ benchmark: the more cores, the lower efficiency. For $HPCG$ benchmark: 
	the "roofline"~\cite{WilliamsRoofline:2009} of that communication intensity was already reached,
	all computers have about the same efficiency.

	According to Eq.~(\ref{eq:soverk}) the efficiency can be interpreted
	in terms of $\alpha$ and $N$,
	and the payload performance of a parallelized sequential computing system can be calculated as
	
	\begin{equation}
	P(N,\alpha) = \frac{N\cdot P_{single}}{{N\cdot \left(1-\alpha\right)+\alpha}}\label{eq:Ppayload}
	\end{equation}
	
\begin{figure*}
		\hspace{-1cm}
		\includegraphics[width=1.15\textwidth]{%fig/
			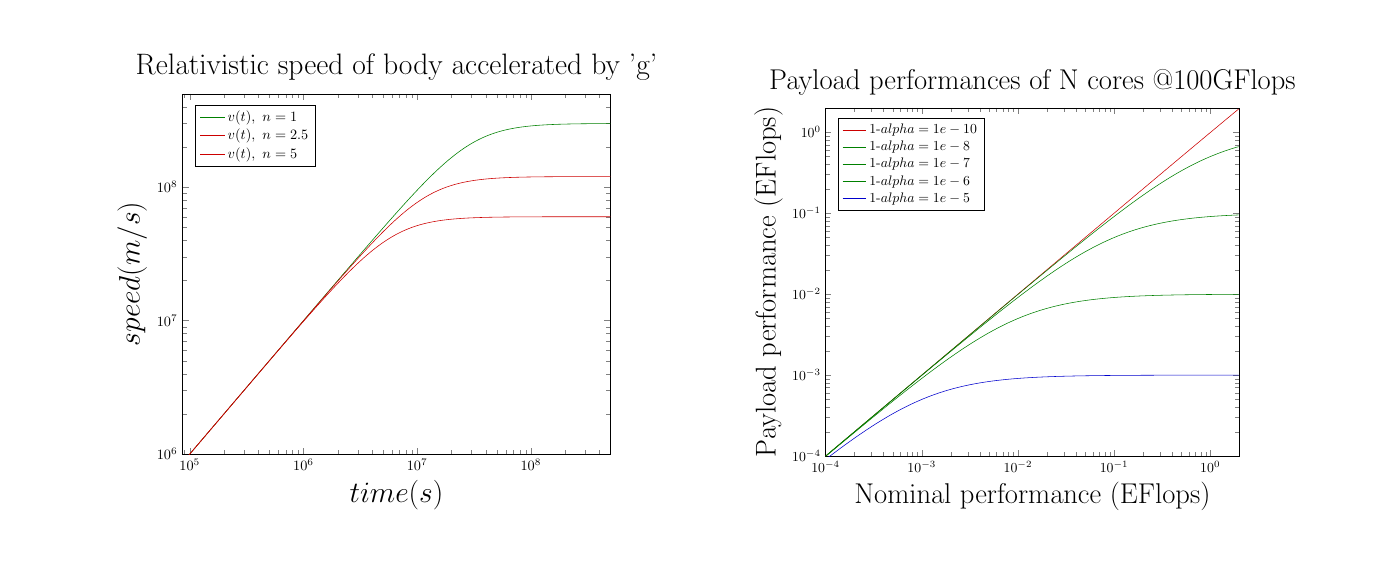}	
		\caption{The analogy between how the speed of light changes the linearity of speed dependence at extremely large speed values and that how the value of performance changes the linearity of performance dependence at extremely large performance values.\label{fig:RelativisticVsPerformance}}
	\end{figure*}

	\noindent This simple formula explains why \textit{the payload performance 
		is not a linear function of the nominal performance} and why
	in the case of very good parallelization ($(1-\alpha)\ll1$)
	and low $N$, one cannot notice the nonlinearity.
	The functional form of the dependence discovers a surprising analogy 
	shown in details in Table~\ref{tab:AnalogyRelativisticPerformance} and Fig~\ref{fig:RelativisticVsPerformance}.

	\begin{table}	
		\caption{The analogy of adding speeds in physics and adding performances in computing, in the classic and modern paradigm, respectively. In both cases a correction term is introduced, that provides noticeable effect only at extremely large values.\label{tab:AnalogyRelativisticPerformance}}
		\maxsizebox{\columnwidth}{!}
		{	
			\begin{tabular}{|p{150pt}|p{165pt}|}
				\hline
				\hline
				Physics & Computing\\
				\hline
				Adding of speeds &	Adding of performance\\
				\hline
				\textcolor{blue}{Classic} & \textcolor{blue}{Classic} \\
				{\large $ v(t) = \textcolor{blue}{t\cdot a}$}
				&
				{\large $ Perf_{total}(N) = \textcolor{blue}{N\cdot P_{single}}$}	\\
				\hline
				t = time &  N = number of cores \\
				\hline
				a = acceleration & $P_{single} = $ Single-performance\\
				\hline
				n = optical density
				&	\textcolor{red}{communication}\\
				\hline
				c = Light Speed &
				\textcolor{red}{$\alpha$ = parallelism}\\
				\hline
				
				\textcolor{red}{ Modern (relativistic)} &	{\textcolor{red}{Modern (Amdahl-aware)~\cite{VeghModernParadigm:2019}, see Eq.~(\ref{eq:soverk})}}  \\
				\vspace{.1cm}
				\Large $ v(t) = \frac{t\cdot a}{\boxed{\textcolor{red}{\sqrt{1+\left(\frac{t\cdot a}{c/n}\right)^2}}}}$
				\vspace{.1cm}
				
				&
				\vspace{.1cm}
				\Large $ P(N) = \frac{N\cdot P_{single}}{\boxed{\textcolor{red}{N\cdot \left(1-\alpha\right)+\alpha}}}$	
				\\
				\hline
				\hline
			\end{tabular}
		}
	\end{table}

	The right side of Fig.~\ref{fig:RelativisticVsPerformance} reveals
	why  the nonlinearity of the dependence of the payload performance on the nominal performance was not noticeable earlier:
	in the age of $1K$ processors, the effect was thousands of times smaller than in the age of $1M$ processors, and the increase seemed to be linear. But anyhow: although
	\textit{one can use the 'weak scaling'  up to around up to a few PFlops (and low communication intensity), it is surely
		not valid any more}. How much the nonlinearity manifests,
	depends on the type of the workload of the computing system~\cite{VeghHowMany:2020}.
	That is, according   Eq.~(\ref{eq:Ppayload}) defines the 'real scaling'.
	\textit{The linear approximation} (according to the 'weak scaling')
	\textit{is not valid any more}, although it \textit{was}  a good approximation at lower performance values and for 
	shorter extrapolation distances.
	
	Notice that in this section, we assumed that $\alpha$ does not depend on $N$. This assumption is undoubtedly valid for low
	number of processors, and unquestionably not valid for the cutting-edge
	supercomputers. 
	That is, as discussed below, the bad news is that the increase of the payload performance is not linear in the function of nominal performance
	(as would be expected based on 'weak scaling'),
	but has a performance limit at which it saturates (according to the first-order approximation) or starts to decrease (according to the second-order approximation). 
	The parallelized sequential processing has different rules of game~\cite{ScalingParallel:1993},~\textbf{\cite{VeghModernParadigm:2019}}: the performance gain ("the speedup") has its inherent bounds~\textbf{\cite{VeghHowMany:2020}}. 
	
	\section{A non-technical model of parallelized sequential operation}
	\label{sec:TheModel}
	
	To understand why the different
	'scaling' methods are approximations with a limited range of validity; we set up a simple non-technical model.

	\begin{figure*}
		\hspace{-0.5cm}\includegraphics[width=\textwidth]{%fig/
			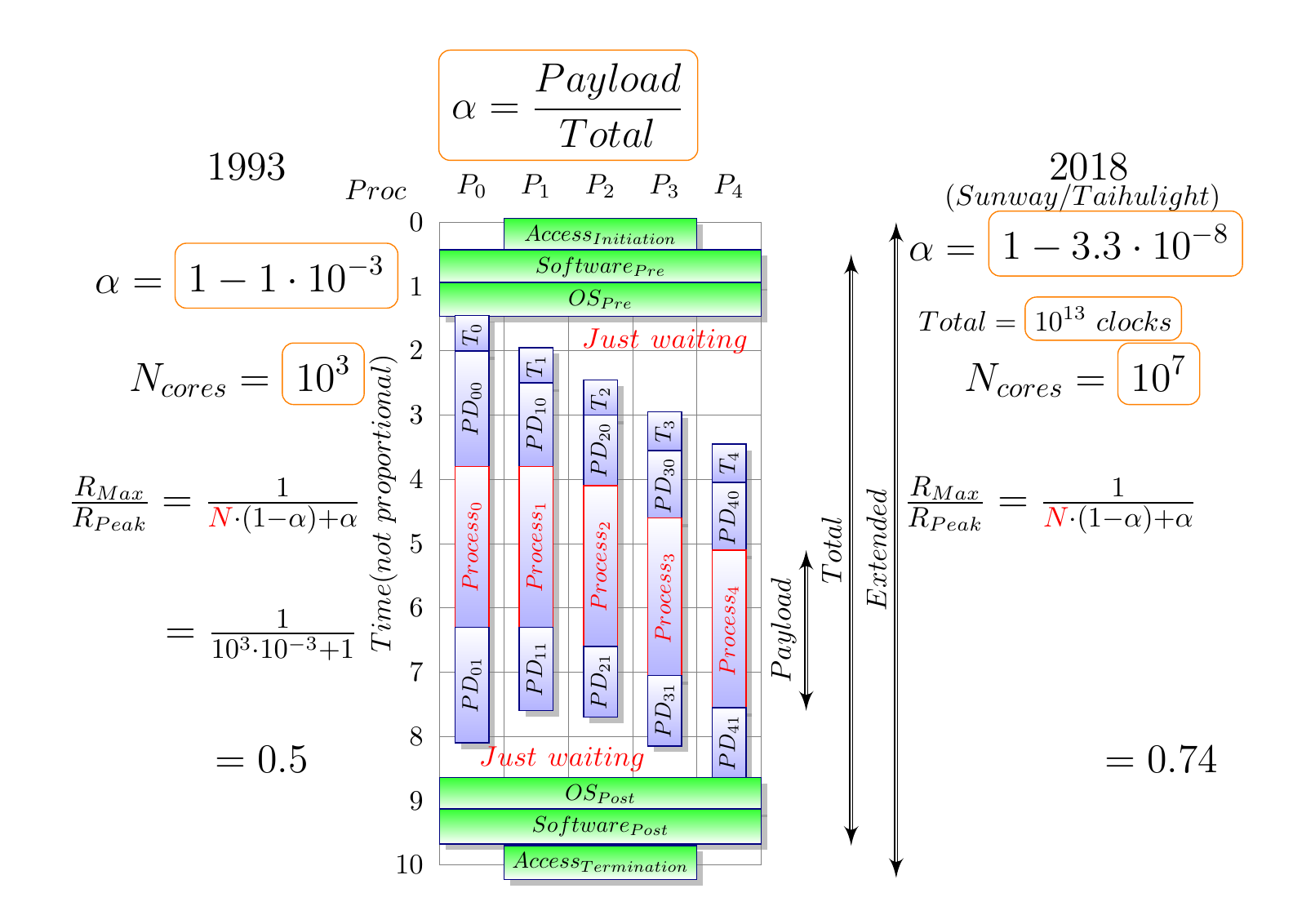}
		
		\caption{A non-technical, simplified model of parallelized 
			sequential computing operations. The contributions of the model component $XXX$ to 
			$\alpha$ (sometimes used as $\alpha_{eff}$ to emphasize that
			it is an effective, empirical value)
			will be denoted by $\alpha_{eff}^{XXX}$ in the text.
			Notice the different nature of those contributions.
			They have only one common feature: \textit{they all consume time}.
			The  vertical scale displays the actual activity for processing units shown on the horizontal scale. 
			\label{fig:AmdahlModelAnnotated}}
	\end{figure*}

	The speedup measurements are simple time measurements\footnote{Sometimes also secondary merits, such as GFlops/Watt or GFlops/USD are also derived} (although they need careful handling and proper interpretation, see good textbooks such as~\cite{RISCVarchitecture:2017}): a benchmark executes a standardized set of machine instructions
	(a large number of times), and divides the known number of operations by the measurement time;
	for both the single-processor and the distributed parallelized sequential systems.
	In the latter case, however, the joint work must also be organized, implemented with additional machine instructions and additional execution time, forming an overhead. This additional activity is \textit{the origin of the inherent efficiency of the parallelized sequential systems}: one of the processors orchestrates the joint operation, the others are idle waiting.
	At this point, the "\textit{dark performance}" appears: the processing units are ready to operate, consume power, but do not make any payload work.
	
	A closer analysis revealed that \textit{one of the essential prerequisites to applying Amdahl's Law is not strictly fulfilled even by the Amdahl's Law}. The reason is that "\textit{It requires the serial
		algorithm to retain its structure such that the same number of instructions are processed by both the serial
		and the parallel implementations for the same input}"~\cite{AmdalVsGustafson96}. Because of this, 
	Amdahl's Law itself is an approximation.
	In its original form, we can call it the \textit{first order approximation} to Amdahl's Law.  The approximation takes that compared to the non-parallelizable payload work, organizing the joint work is negligible. The validity of this assumption
	is limited to a very low number of cores and a relatively high ratio of overhead. Recall that in the age of Amdahl, the non-payload workload ratio was in the range of dozens of percent, so some extra work did not make a considerable difference. Today, as we discuss below,
	\textit{the ratio of the overhead is by orders of magnitude lower, while the number of cores is by orders of magnitude higher} (see also the parameters of the different configuration in Fig.~\ref{fig:AmdahlModelAnnotated}). This latter aspect we consider
	as the \textit{second order approximation} to Amdahl's Law; for more details see~\cite{VeghHowMany:2020}.

	Amdahl's idea is to \textit{put everything that we cannot parallelize, i.e., distribute between the fellow processing units, into the sequential-only fraction}. In the spirit of this idea, for describing the parallel operation
	of sequentially working units, we prepared the model depicted in Figure~\ref{fig:AmdahlModelAnnotated}.
	The technical implementations of the different parallelization methods show up virtually infinite variety~\cite{HwangParallelism:2016}, so we present here a (by intention) strongly simplified model.
	The non-parallelizable contributions are 
	virtually classified (sometimes contracted)
	and shown as general contribution terms in the figure. 
	In this way, the model is general enough to discuss some case studies of parallelly working systems qualitatively, neglecting different contributions
	as possible. One can convert the model to a technical (quantitative) one via interpreting the contributions in technical terms, although with some obvious limitations.

	As Figure~\ref{fig:AmdahlModelAnnotated}
	shows, in the parallel operating mode (in addition  to the calculation, furthermore the communication of data between the processing units) \textit{both the software
		(in this sense: computation and communication, including data access)
		and the 
		hardware (interconnection, accelerator latency) contribute to the execution time}, i.e., one must consider both of those components in Amdahl's Law. This statement is not new, again: see
	\cite{AmdahlSingleProcessor67,AmdalVsGustafson96}.
	
	The non-parallelizable (i.e. apparently sequential) part comprises contributions from \gls{HW}, \gls{OS}, \gls{SW} and \gls{PD} (the "\textit{propagation rates of different physical effects}"), and also some access time is needed  for reaching the parallelized system.
	This separation is rather conceptual than strict, although dedicated measurements can reveal their role, at least approximately.
	Some features can be implemented in either \gls{SW} or \gls{HW}, or shared between them, and also some sequential activities
	may happen partly parallel with each other.
	The relative weights of the contributions are very different for different parallelized systems,
	and even within those cases depend on many specific factors,
	so \textit{in every single parallelization case, it requires a careful analysis}.
	The \gls{SW} activity represents what was assumed by Amdahl as the total sequential fraction.
	What did not yet exist in the age of Amdahl, the non-determinism of the modern \gls{HW} systems~\cite{PerformanceCounter2013}~\textbf{\cite{Molnar:2017:Meas}} also contributes to the non-parallelizable portion of the task: the slowest unit defines the resulting execution time
	of the parallelly working processing elements. Also, notice that optimization possibilities are present in the system; for example, see in Fig.~\ref{fig:AmdahlModelAnnotated} how the contribution of classes propagation delay and looping delay can be combined to achieve a better timing.
	
	Our model assumes no interaction between the processes
	running on the parallelized systems in addition to the necessary minimum: starting and terminating the otherwise independent processes, which take parameters at the beginning and return a result at the end.
	It can, however, be trivially extended to the more general case when processes must share some resources (like a database, which shall provide different records for the different processes), 
	either implicitly or explicitly. The concurrent objects have their inherent sequentiality~\cite{InherentSequentiality:2012}, and the synchronization and
	communication between those objects considerably increase~\cite{YavitsMulticoreAmdahl2014} the non-parallelizable portion
	(i.e., contribution to $(1-\alpha_{eff}^{SW})$ or  $(1-\alpha_{eff}^{OS})$), so in the case of the massive number of processors, special attention must be devoted to their role on the efficiency of the application on the parallelized system.

	In the case of distributed systems, the physical size of the computing system also matters:
	the processor, connected to the first one with a cable of length of dozens of meters,  must spend several hundred clock cycles with waiting,
	only because of the finite speed of propagation of light, topped by the latency time and hoppings of the interconnection (not mentioning geographically distributed computer systems, such as some clouds, connected through general-purpose networks).
	\textit{This aspect is completely neglected in the 'weak scaling' approximation.}
	Detailed calculations are given in~\textbf{\cite{Vegh:2017:AlphaEff}}.

	After reaching a certain number of processors, there is no more increase in the payload fraction when adding more processors:
	the first fellow processor already finished the task and is idle waiting, while the last one is still idle waiting for the start command. One can increase this limiting number by organizing the processors into clusters; then, the first computer must speak \textit{directly} only to the head of the cluster.
	Another way is to  distribute the job near to the processing units,
	either inside the processor~\cite{CooperativeComputing2015} 
	or using processors to let do the job by the processing units
	of a GPGPU.

	This looping contribution is not considerable (and so: not noticeable) at a low number
	of processing units, but can be a dominating factor at a high
	number of processing units. This "high number" was a few dozens
	at the time of writing the paper~\cite{ScalingParallel:1993}, today
	it may be in the order of a few millions.
	Considering the effect of the looping contribution is the borderline between the first and second-order approximations in modeling the performance: the housekeeping keeps growing with
	the growing number of processors, while the resulting performance
	does not increase anymore. Even, the housekeeping gradually becomes the dominating factor of the performance limitation, and leads to a decrease in the payload performance: "\textit{there comes a point when using more processors \dots increases the execution time rather than reducing it}"~\cite{ScalingParallel:1993}.  That is, the first-order approximation results only in saturated performance; the second-order approximation leads to reaching an inflection point followed by
	decreasing performance and efficiency.

	\section{Application fields\label{sec:ApplicationFields}}
	
	According to the model, one expects $(1-\alpha_{eff})$  to describe the fraction of the total
	(even unintended or only apparently) sequential part in any \gls{HW}/\gls{SW} system, and
	\textit{it is a sensitive measure of
		disturbances and inefficiencies of parallelization}~\cite{Vegh:2017:AlphaEff}. 
	This value can be used as the merit to compare setups, 
	computers manufactured in different ages with different technologies, conditions of network operation, the algorithm for
	communication within a closed chip, \gls{SW} load balancing.
	
	In this section (except section~\ref{sec:AccessTime}) we assume that  
	the parallelized computing system is accessible in a negligible time, and that the parallelized system under study is properly defined.
	We do not care whether the one-time contributions (such as initiating the data structures and starting the calculations)
	are done by the user \gls{SW} or by the \gls{OS}; furthermore
	we assume that we repeat the payload calculation so many times that
	we can neglect the one-time contributions.

	\begin{figure*}[b!th]
		\maxsizebox{.9\textwidth}{!}
		{
			\pgfplotsset{width=8.5cm}
			\begin{tabular}{cc}
				\begin{tikzpicture}
				\begin{axis}[
				cycle list name={my color list},
				legend style={
					cells={anchor=east},
					legend pos={north west},
				},
				xmin=0, xmax=9,% x scale
				ymin=0., ymax=0.9, % y scale
				xlabel=Number of processors,
				ylabel=1 - (Speedup/No of processors),
				%		ymode=log,
				]
				\addplot[ solid, color=webblue, every mark/.append style={solid, fill=webblue}, mark=*%mark=x,red
				] plot coordinates {
					(1, 0) %1) 
					(2, 0.026) %.974)    
					(3, 0.315) %.685) 
					(4, 0.597)
					(6, 0.403) %.571) 
					(8, 0.537) %.463) 
				};
				\addlegendentry{Audio stream 1}
				\addplot[ solid, color=webblue, every mark/.append style={solid, fill=webred},mark=diamond*%mark=*,red
				] plot coordinates {
					(1, 0) %1) 
					(2, 0) %1)    
					(3, 0.067) %.933) 
					(4, 0.079) %.921)
					(6, 0.222) %.778) 
					(8, 0.294) %.706) 
				};
				\addlegendentry{Audio stream 2}
				
				\addplot[dashed, color=webred, every mark/.append style={solid, fill=webblue}, mark=* %mark=x,blue,thick
				] plot coordinates {
					(1, 0) %1) 
					(2, .149) %.851)    
					(4, .444) %.556)
					(8, .722) %.278) 
				};
				\addlegendentry{Radar initial}
				\addplot[ dashed, color=webred, every mark/.append style={solid, fill=webred}, mark=diamond*%
				%mark=*,blue,thick
				] plot coordinates {
					(1, 0) %1) 
					(2, .119) %.881)    
					(4, .266) %.734)
					(8, .449) %.551) 
				};
				\addlegendentry{Radar improved}

				\end{axis}
				\end{tikzpicture}		
				&
				\begin{tikzpicture}
				\begin{axis}[
				cycle list name={my color list},
				legend style={
					cells={anchor=east},
					legend pos=south east,
				},
				xmin=1, xmax=9,% x scale
				ymin=.0002, ymax=0.4, % y scale
				xlabel=Number of processors,
				ylabel={1-$\alpha_{eff}$},
				ymode=log,
				%		ymode=lin,
				]				
				
				\addplot[ solid, color=webblue, every mark/.append style={fill=webblue}, mark=*%thick, color=red,mark=x
				]
				plot coordinates {
					(2, 0.027) 
					(3, .23) 
					(4, .226) 
					(6, .151) 
					(8, .166) 
				};
				\addlegendentry{Audio stream 1}
				
				\addplot[solid, color=webblue, every mark/.append style={ fill=webred},mark=diamond*%thick, color=red,mark=*
				]
				plot coordinates {
					(2, 0) 
					(3, .036) 
					(4, .029) 
					(6, .057) 
					(8, .060) 
				};
				\addlegendentry{Audio stream 2}
				
				\addplot[dashed, color=webred, every mark/.append style={solid, fill=webblue}, mark=*%thick, color=blue,mark=x
				]
				plot coordinates {
					(2, 0.174) 
					(4, .266) 
					(8, .371) 
				};
				\addlegendentry{Radar initial}
				\addplot[ dashed, color=webred, every mark/.append style={solid, fill=webred}, mark=diamond*%
				%thick, color=blue,mark=*
				]
				plot coordinates {
					(2, 0.135) 
					(4, .121) 
					(8, .117) 
				};
				\addlegendentry{Radar improved}

				\end{axis}
				\end{tikzpicture}
				\\
			\end{tabular}
		}
		\caption{Relative speedup (left side)
			and ($1-\alpha_{eff}$) (right side) values, measured running the audio and radar processing on
			different number of cores. \protect{\cite{CompilerInfrastructure:2014}}\label{fig:CompilerDiagram}}	
	\end{figure*}
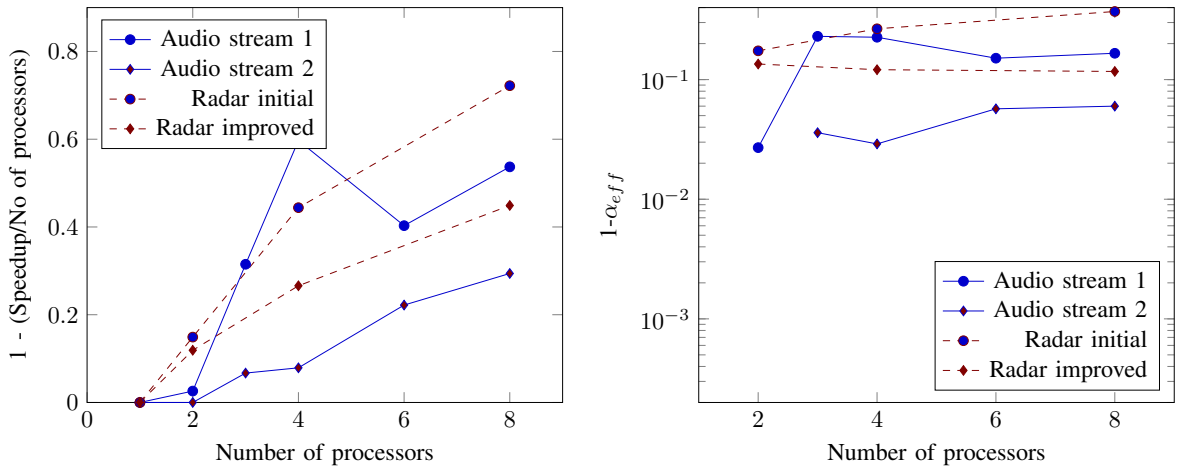

	\subsection{Load balancing compiler\label{fig:LoadBalancing}}
	
	Today, mainly because of the more and more widespread utilization of multi-core processors,
	more and more applications are considered to be re-implemented in multi-core aware form.
	Because it is a serious (and expensive!) effort, before deciding to start such a re-implementation,
	one needs to guess
	the speed gain. After finishing the re-implementation,
	it would be desirable to measure whether we achieved the goal. A method to find out during development, whether we can increase further parallelization
	using a reasonable amount of development work, would be highly desirable.
	The achievable speed gain depends on both the structure of the code and the hardware architecture,
	so one must scrutinize all those aspects.
	
	A compiler making load balancing 
	of an originally sequential code for different number of cores
	is described and validated  in paper~\cite{CompilerInfrastructure:2014},
	by running the executable code on platforms having different numbers of cores.
	In terms of efficiency, the results they presented have common features
	and can be discussed together.
	
	The left subfigure of Fig.~\ref{fig:CompilerDiagram} (Fig 8 in~\cite{CompilerInfrastructure:2014}) displays their results
	in function of the number of cores,
	using the figure of merit the authors used, the efficiency $E$ (see Equ. (\ref{eq:soverk})).
	The data displayed in the figures are derived simply through reading back
	diagram values from the mentioned figures in~\cite{CompilerInfrastructure:2014},
	so they may not be accurate. However, they are accurate enough to support our conclusions.

	Their first example shows the results of implementing parallelized processing of an audio stream manually,
	with an initial (first attempt), and more careful (having already experienced programmers) implementation.
	For the two different processings of audio streams,
	using efficiency $E$ as merit enables only to claim
	a qualitative statement about load balancing, that "\textit{The higher number of parallel processes in Audio-2 gives better results}", because the Audio-2 diagram decreases less steeply than Audio-1.
	In the first implementation, where the programmer had no previous experience with
	parallelization, the efficiency quickly drops with the increasing number of cores.
	In the second round, with experiences from the first implementation, the loss is much less,
	so $1-E$ rises less speedily.
	
	Their second example is processing radar signals.
	Without switching in their compiler the load balancing optimization on, the slope of the $1-E$ diagram line is much more significant.
	It seems to be unavoidable that as the number of cores increases, the efficiency 
	(according to Eq. (\ref{eq:soverk}))  decreases, even at such a low number of cores.
	Both examples leave
	the questions open whether further improvements are possible
	or whether the parallelization is uniform in function of the number of cores.
	
	In the right subfigure of Fig.~\ref{fig:CompilerDiagram} (Fig. 10 in~\cite{CompilerInfrastructure:2014}) the diagrams show the $(1-\alpha_{eff})$ values,
	derived from the same data.
	In contrast with the left side, these values are nearly constant 
	(at least within the measurement data readback error) which means that 
	the derived merit value is characteristic of the system.
	By recalling Equ. (\ref{eq:AmdahlBase}) one can identify this parameter
	as the resulting non-parallelizable part of the activity,
	which -- even with careful balancing -- one cannot distribute between the cores,
	and cannot be reduced.
	
	In the light of this, one can conclude that both the programmer in the case of 
	audio stream and the compiler in the case of radar signals
	correctly identified and reduced the amount of non-parallelizable activity.
	The $\alpha_{eff}$ is practically constant in function of the number of cores,
	nearly all optimization possibilities found and they hit the wall due to 
	the unavoidable contribution of non-parallelizable software contributions.
	The better parallelization leads to lower 
	$(1-\alpha_{eff})$ values, and less scatter in the function of the number of cores.
	The uniformity of the values also make highly probable, that in the case of audio
	streams further optimization can be done, at least for 6-core and 8-core systems,
	while the processing of radar signals reached its bounds.
	
	Note that we must not compare the absolute values for analyzing different programs:
	they represent the sequential-only part of two programs,
	which may be different.
	It looks like that the $(1-\alpha_{eff})$ imperfectness can be reduced
	to about $10^{-1}$ with software methods of parallelization.

	\begin{figure*}
		\includegraphics[width=\textwidth]{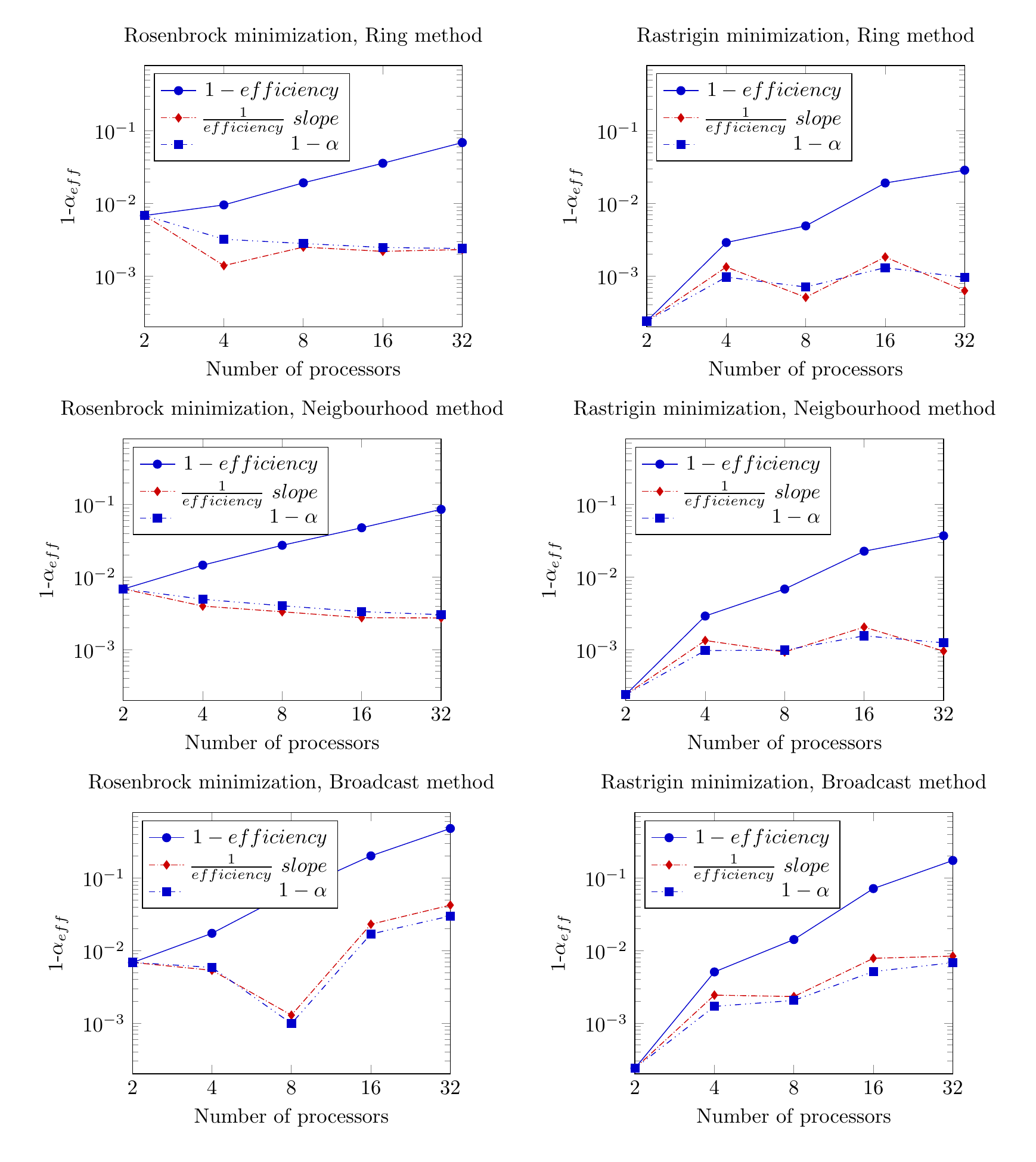}	
		\caption{Comparing efficiency, efficiency slope and $\alpha_{eff}$
			for different communication strategies when running two minimization task on SoC by~\cite{ReconfigurableAdaptive2016}
			\label{fig:SoCCommunication}}
	\end{figure*}
	
	\subsection{Measuring the efficiency of the on-chip networking\label{sec:AccessTime}}
	
	It is not a trivial task to find out the subtle points
	of on-chip networking, because of both the limited
	accessibility and the low number of processing units.
	The merit developed here, however, can also help in that case;
	although the available non-dedicated measurements enable us to draw only conclusions of limited accuracy.
	
	In~\cite{ReconfigurableAdaptive2016} the authors compare
	different communication strategies  their \gls{PSO} uses when minimizing  
	Rosenbrock's function and Rastrigin's functions, respectively.
	%In this case, the access time can be neglected, and the number of cores is low enough to neglect the linear contributions.
	From their data, we calculated the corresponding $\alpha_{eff}$ values and displayed them in Fig.~\ref{fig:SoCCommunication}. 
	The fluctuations seen in the figure
	show the limitations of the (otherwise excellent) measurement precision;
	for this type of investigation, one would need much longer measurement time.
	\index{alpha@$\alpha_{eff}$!communication strategies}

	\begin{figure}
		\includegraphics[width=\columnwidth]{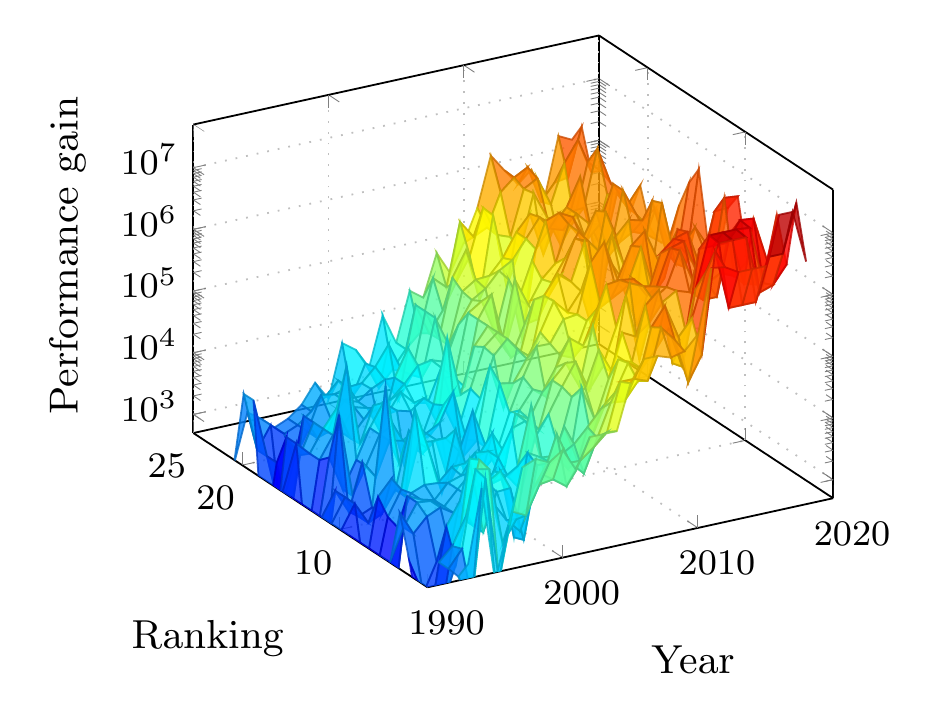}	
		\caption{The history of supercomputing in terms of performance gain \label{fig:SupercomputerGainHillside}}
	\end{figure}

	The contribution of the \gls{OS} cannot be separated, again, from \gls{SW} contribution. Although the precision of the available data does not enable to make a detailed analysis of the behavior of the scaling and to qualify the communication method exhaustively,
	some observations we can make.
	When utilizing only two cores, the variety of ways to communicate is minimal. In this case, all communication methods delivered the same $\alpha_{eff}$ value, proving the self-consistency of both the model and the measurement.
	The values of $\alpha_{eff}$ ($5*10^{-3}$ and $2*10^{-4}$) deviate considerably for the two minimization methods, however. We attribute this deviation to the different structures ($\alpha_{eff}^{SW}$) of the two applications.
	As the diagrams of the Rastrigin method show,
	the contribution due to the propagation delay can be in the order of $1*10^{-3}$; which is considerable for the Rastrigin method, but not for the Rosenbrock method.
	The reason why for higher core numbers is that  $\alpha_{eff}$ is nearly constant in the case of the first two communication methods: one of the contributions dominates; although for the Rastrigin method $\alpha_{eff}^{SW}$, while for the Rastrigin method $\alpha_{eff}^{communication}$ is the dominating term.
	
	A bit different is the case for the broadcast-type communication, for both types of minimization methods: the resulting $(1-\alpha_{eff})$ increases with the increasing number of cores.
	Here the reason is that the number of collisions (and so the time spent with waiting for repeating) increases with the number of cores. 
	This contribution increasingly dominates for the Rastrigin case
	and increases moderately the already high $(1-\alpha_{eff})$ value at a high number of cores, while at a low number of cores the $(1-\alpha_{eff}^{SW})$ value persists in dominating for the Rosenbrock case.
	
	\begin{figure*}
		\maxsizebox{\textwidth}{!}
		{
			\begin{tabular}{cc}
				\includegraphics[scale=1.1]{%fig/
					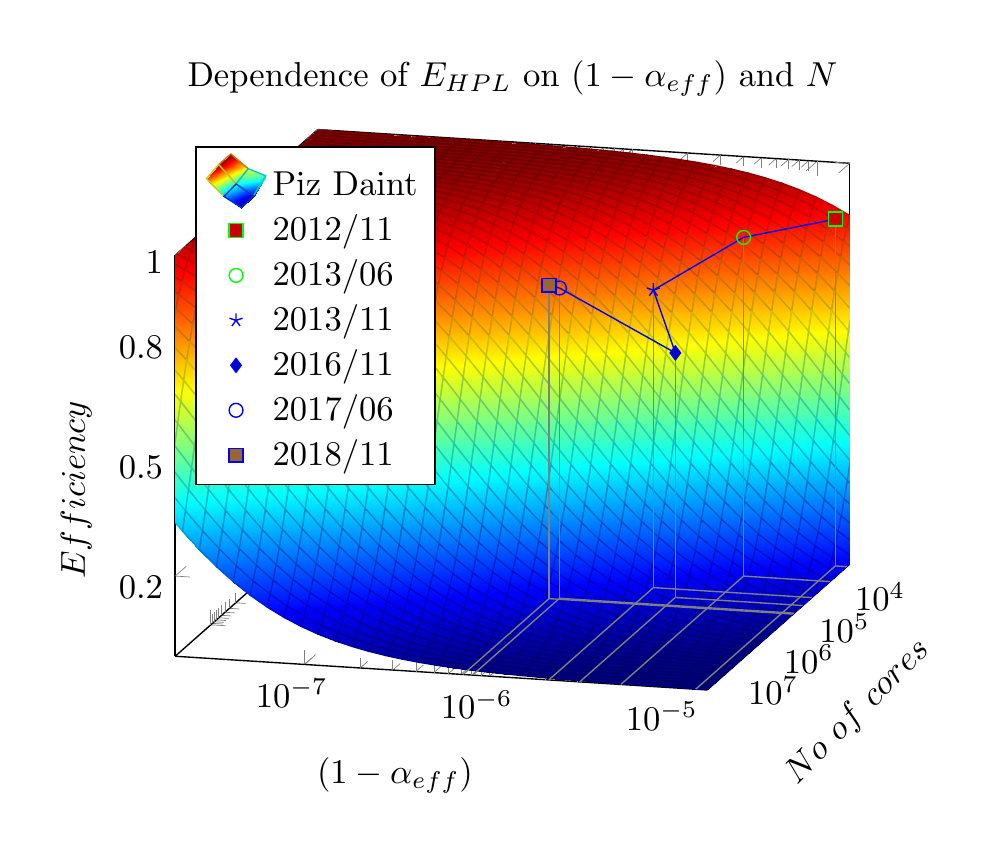}
				&
				\includegraphics[scale=.785]{%fig/
					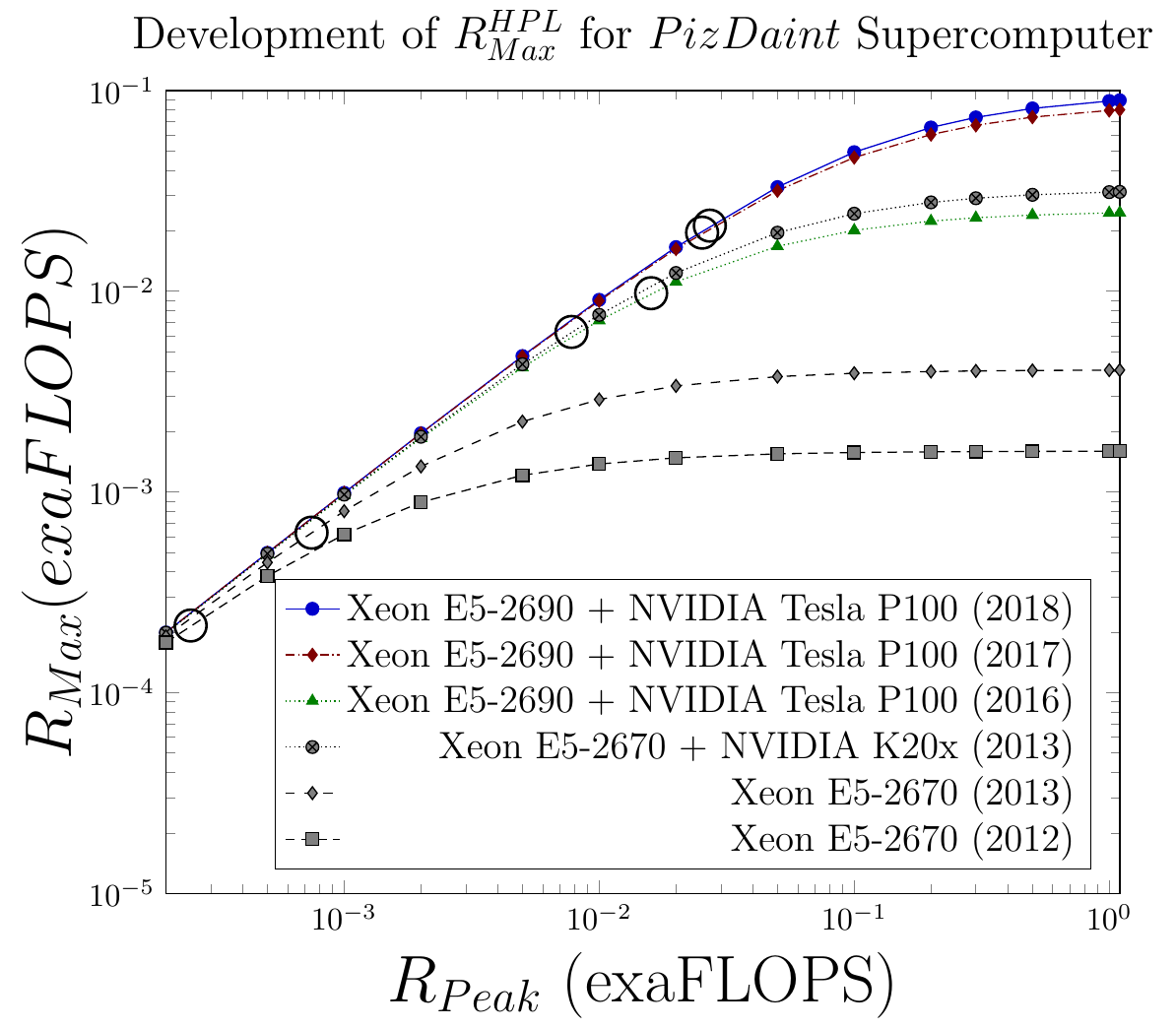}
			\end{tabular}
		}
		\caption{The history of supercomputer Piz Daint in terms of  efficiency and  payload performance. 
			The left subfigure shows how the efficiency changed
			as the developers proceeded towards higher performance. The right subfigure shows the reported 
			performance data (the bubbles), together with the 
			diagram line calculated from the value as described above. Compare the value ot the diagram line
			to the measured performance data in the next reported stage. 
			\label{fig:PizDaintEfficiency}
		}	
	\end{figure*}

	\subsection{The history of supercomputing}
	The TOP500 database~\cite{TOP500:2017}
	provides all needed data to calculate $\alpha$,
	independently from the date of manufacturing, technology,
	manufacturer, number, and kind of processors. We can use the
	parallelization efficiency in studying (among others) the 
	history of supercomputing.

	During the past quarter of a century, the proportion of the contributions changed considerably: 
	today the number of processors is thousands of 
	times higher than it was a quarter of a century ago. 
	The growing physical size and the higher processing speed
	increased the role of the propagation overhead,
	furthermore the large number of processing units strongly amplified 
	the role of the looping overhead.
	As a result of the technical development, \textit{the phenomenon 
		on the performance limitation~\cite{ScalingParallel:1993} returned in a technically 
		different form at a much higher number of processors}.
	\noindent 
	As discussed, except for an extremely high number of processors, it can be assumed  that $\alpha$ is independent
	from the number of processors. 
	Equ.~(\ref{eq:alphafromr}) can be used to derive quickly the value of $\alpha$ from the values of parameters $R_{Max}/R_{Peak}$ and the number of cores  $N$.

	\subsection{The effect technology change in supercomputing\label{sec:AccessTime}}
	
	As expressed by Eq.~(\ref{eq:Ppayload}), the resulting performance of parallelized computing systems depends on both the single-processor performance and performance gain (mainly the perfectness of the parallelization).
	To separate these two factors, 
	Fig.~\ref{fig:SupercomputerGainHillside} displays \textit{the performance gain} of the supercomputers in the function of their year of construction and the ranking in the given year. The "hillside" reflects the enormous development of the parallelization technology. Unfortunately, the different individual factors (such as interconnection quality, using accelerators and clustering, or using slightly different computing paradigm) cannot be separated in this way, although even in this figure some limited validity conclusions can be drawn.
	
	One can localize two 'plateaus' before the year 2000 and after the year 2015, unfortunately, underpinning Amdahl's Law and refuting Gustafson's Law.
	The values between 2000 and 2010 demonstrate the development of the interconnection technology
	(for a more detailed analysis, see Fig.~6 in~\cite{VeghHowMany:2020}). Before 2010 running the benchmark on a top supercomputer was a communication-bound task. Since 2015 it is a computing-bound task.
	The appearance of accelerators 
	around 2011 caused some 'humps', and both the excellent clustering and using 'cooperating cores'~\cite{CooperativeComputing2015} increased the achieved performance gain. The wide variety of technical components, manufacturers, interconnections, processors, accelerators,  does not enable us to make more detailed connections.

	Fortunately, both the validity of the 'real scaling' and the accuracy of the merit in predicting the future performance values 
	can be demonstrated.
	The supercomputers usually do not have a long lifespan and several documented stages. 
	One of the rare exceptions is the supercomputer \textit{Piz Daint}.
	The documented lifetime spans over 6 years, and during
	that time, a different number of cores, without and with acceleration, using different accelerators, were used.

	\begin{figure*}
		\includegraphics[width=\textwidth]{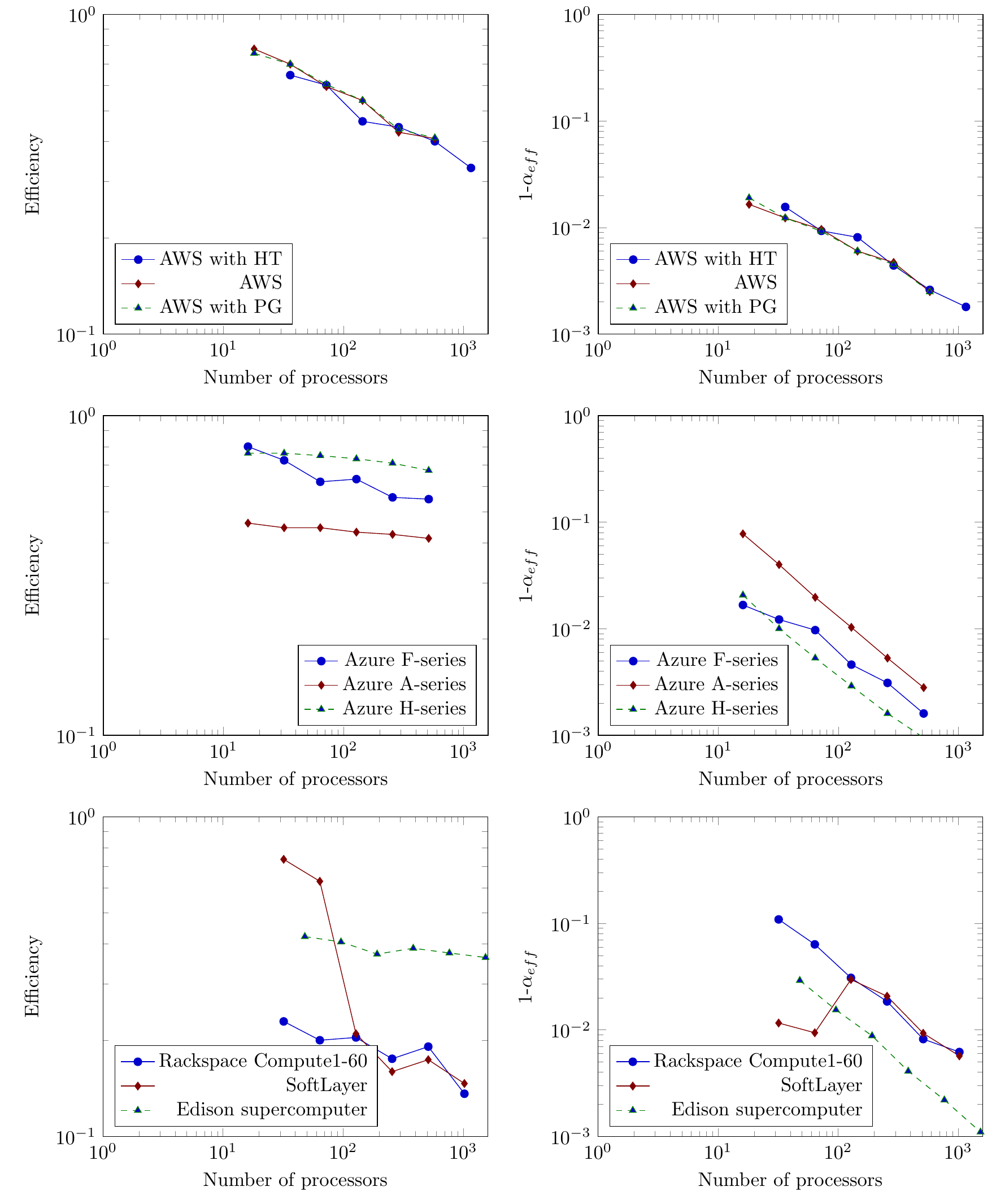}	
		\caption{The effect of neglecting the access time when measuring efficiency of some cloud services\label{fig:CloudEfficiency}}
	\end{figure*}

	Figure~\ref{fig:PizDaintEfficiency} depicts the performance and efficiency values
	published during its lifetime, together with the diagram lines predicting (at the time of making the prediction) the values
	at the higher nominal performance values.
	The left subfigure shows how the changes made in the configuration affected the efficiency (the timeline starts in the top right corner, and a line connects the consecutive stages). 
	
	In the right subfigure, the bubbles represent 
	data published in the adjacent editions of the TOP500 lists,
	the diagram lines crossing them  are the predictions
	made from that snapshot.
	We can compare the predicted values to the value
	published in the next list.
	It is especially remarkable that 
	introducing GPGPU acceleration resulted only in a slight increase (in good agreement with~\cite{Lee:GPUvsCPU2010}
	and~\cite{WhyNotExascale:2014})
	compared to the value expected based purely on the increase of the number of cores. Between the samplings more than one parameter was changed,
	that is, we cannot demonstrate the net effect of a change.
	However, the measured data sufficiently underpin our limited validity conclusions and show that the theory correctly describes
	the tendency of the development of the performance and the efficiency. The predicted performance values are also reasonably accurate.
	
	Introducing a GPU accelerator is a one-time performance increase step~\cite{WhyNotExascale:2014}, and cannot be taken into account by the theory.
	Notice that introducing the accelerator increased the payload performance, but decreased the efficiency (copying data from one address space to another increases latency). 
	Changing the accelerator to another type with slightly higher performance (but higher latency due to the larger GPGPU memory) caused a slight \textit{decrease} in the efficiency.
	
	\subsection{The effect of not considering the access time\label{sec:AccessTime}}
	
	In~\cite{BenchmarkingClouds:2017} the authors benchmarked some commercially available cloud services, fortunately using \gls{HPL} benchmark.
	Fig.~\ref{fig:CloudEfficiency} shows 
	on the left side the efficiency (i.e. $\frac{R_{Max}}{R_Peak}$),
	on the right side, the $(1-\alpha)$ values, in the function of the number of processors in the used configuration.
	One can immediately notice on one side
	that the values of  $\frac{R_{Max}}{R_Peak}$ are considerably lower than unity,
	even for a very low number of cores. On the other side, the $(1-\alpha)$ values steeply \textit{decrease} as the number of cores \textit{increases}, although the model contains only contributions which may only \textit{increase} as the number of cores increases.
	
	The benchmark \gls{HPL} characterizes 
	the setup, so the benchmark is chosen correctly.
	When acquiring measurement data, in the case of clouds,
	the access time must also be considered, see Fig.~\ref{fig:AmdahlModelAnnotated}.
	If one measures the time on the client's computer (and this is what is possible using those services),
	one uses the time \textit{Extended} in the calculation in place of \textit{Total}. That is, the 'device under test' is chosen improperly.
	
	This artifact is responsible for both mentioned differences.
	The efficiency measured in this way would not achieve 100~\% even on a system comprising only one single processor.
	Since $\alpha$ measures the \textit{average} utilization of processors, this foreign contribution is divided by the number of processors, so with increasing the number of processors, the relative weight of this foreign contribution decreases, causing to decrease the calculated value of $(1-\alpha)$.
	Since the access is provided through the Internet where
	the operation is stochastic, the measurements cannot be
	as accurate as in purpose-built systems\footnote{
		A long term systematic study~\cite{PricePerformanceClouds:2017}
		derived the results that measured data show dozens of percent of the variation in long term run, and also unexpected variation in short term run.}.
	Some qualitative conclusions of limited validity,
	however, can be drawn even from those data.
	
	\textit{At such a low number of processors} neither of the contributions
	depending on the processor number is considerable, so one can
	expect that in the case of correct measurement  $(1-\alpha)$ would be constant.
	So, extrapolating the diagram lines of $(1-\alpha)$ to the value corresponding to a one-processor system, one can see that
	both for Edison supercomputer and Azure A series  grid (and maybe also Rackspace)
	the expected value is approaching unity (but obviously below it).
	From the slope of the curve (increasing the denominator 1000 times, $(1-\alpha)$ reduces to $10^{-3}$), and one can even find out that $(1-\alpha)$ should be around  $10^{-3}$.
	Based on these data, one can agree with the conclusion that 
	--on a good grid-- benchmark \gls{HPCG} can run as effectively as on the supercomputer they used. 
	One should note, however, that  $(1-\alpha)$ is about 3 orders of magnitude better for TOP500 class supercomputers,
	but this makes a difference only for \gls{HPL} class benchmarks
	and only at a large number of processors.
	This conclusion can be misleading: \textit{whether a high-performance cloud can replace a supercomputer in solving a task, strongly depends on the number of cores}, because of the different $\alpha$ values.
	
	Note that in the case of AWS grids and Azure F series 
	the $\alpha_{eff}^{OS+SW}$ starts at about $10^{-1}$,
	and this is reflected by the fact that their efficiency drops quickly as the number of the cores increases.
	Interesting to note that ranking based on  $\alpha$ is
	just the opposite of the ranking based on efficiency (and strongly correlates with the price of the service).
	
	One can also extrapolate the efficiency values to the point
	corresponding to one core only. In the case of measurement with no such artifact, the backprojected\textit{ efficiency} value should be
	around unity. If the measurement artifact is present,
	it is not the case; the values deviate by a factor up to 2.
	The back-projected $(1-\alpha)$ values are much more consistent:
	they tend to hit the value of unity. Sometimes it is lower
	(meaning some other, foreign,  performance loss),
	but in no case higher than unity.

	\section{Conclusion}
	The scaling methods, mainly due to their simplicity, can be useful when applied in the range of their validity. Given that they are approximations,
	we must scrutinize the validity of the omissions periodically. The approximations to the performance of
	parallelized sequential systems routinely deployed the
	'weak scaling' method to estimate the payload
	performance of future, ever-larger scale system;
	without scrutinizing the validity of the method
	under the current technical situation.
	However, using this approximation (the incremental development) led to 
	unexpected phenomena, failed supercomputers
	and unexpectedly low efficiency of the systems.
	The 'real scaling' is in complete agreement with the
	experiences and measured values.

	%\bibliographystyle{IEEEtran}
	%%% argument is your BibTeX string definitions and bibliography database(s)
	%\bibliography{../../CommonBibliography,../../CommonPrivateBibliography}
	%
	%% biography section
	%% 
	%% If you have an EPS/PDF photo (graphicx package needed) extra braces are
	%% needed around the contents of the optional argument to biography to prevent
	%% the LaTeX parser from getting confused when it sees the complicated
	%% \includegraphics command within an optional argument. (You could create
	%% your custom macro containing the \includegraphics command to make things
	%% simpler here.)
	%%\begin{IEEEbiography}[{\includegraphics[width=1in,height=1.25in,clip,keepaspectratio]{mshell}}]{Michael Shell}
	%% or if you just want to reserve a space for a photo:
	%\end{IEEEbiography}
	%
	%%\newpage
	%%\vfill
	%
	%
	%% that's all folks
\end{document}